\documentclass[twocolumn]{aastex631}

\shorttitle{FRB and PRS Populations}
\shortauthors{Law, Connor, \& Aggarwal}
\graphicspath{{./}{figures/}}

\begin{document}

\title{On the Fast Radio Burst and Persistent Radio Source Populations}

\author[0000-0002-4119-9963]{Casey J.~Law}
\affiliation{Cahill Center for Astronomy and Astrophysics, MC 249-17 California Institute of Technology, Pasadena, CA 91125, USA}
\affiliation{Owens Valley Radio Observatory, California Institute of Technology, 100 Leighton Lane, Big Pine, CA, 93513, USA}

\author{Liam Connor}
\affiliation{Cahill Center for Astronomy and Astrophysics, MC 249-17 California Institute of Technology, Pasadena, CA 91125, USA}

\author[0000-0002-2059-0525]{Kshitij Aggarwal}
\affil{West Virginia University, Department of Physics and Astronomy, P. O. Box 6315, Morgantown, WV, USA}
\affil{Center for Gravitational Waves and Cosmology, West Virginia University, Chestnut Ridge Research Building, Morgantown, WV, USA}

\begin{abstract}
The first Fast Radio Burst (FRB) to be precisely localized was associated with a luminous persistent radio source (PRS). Recently, a second FRB/PRS association was discovered for another repeating source of FRBs. However, it is not clear what makes FRBs or PRS or how they are related. We compile FRB and PRS properties to consider the population of FRB/PRS sources. We suggest a practical definition for PRS as FRB associations with luminosity greater than $10^{29}$\ erg s$^{-1}$\ Hz$^{-1}$ that is not attributed to star-formation activity in the host galaxy. We model the probability distribution of the fraction of FRBs with PRS for repeaters and non-repeaters, showing there is not yet evidence for repeaters to be preferentially associated with PRS. We discuss how FRB/PRS sources may be distinguished by the combination of active repetition and an excess dispersion measure local to the FRB environment. We use CHIME/FRB event statistics to bound the mean per-source repetition rate of FRBs to be between 25 and 440 yr$^{-1}$. We use this to provide a bound on the density of FRB-emitting sources in the local universe of between $2.2\times10^2$\ and $5.2\times10^4$\ Gpc$^{-3}$\ assuming a pulsar-like beam width for FRB emission. This density implies that PRS may comprise as much as 1\% of compact, luminous radio sources detected in the local universe. The cosmic density and phenomenology of PRS are similar to that of the newly-discovered, off-nuclear ``wandering'' AGN. We argue that it is likely that some PRS have already been detected and misidentified as AGN.
\end{abstract}


\section{Introduction}
\label{sec:intro}


Several hundred Fast Radio Burst (FRB) sources have been identified in the last decade \citep{2007Sci...318..777L, 2019A&ARv..27....4P, 2019ARA&A..57..417C, 2021arXiv210604352T}. Recent observational efforts have focused on localizing FRBs to arcsecond precision in order to associate the burst to multiwavelength counterparts \citep{2010PASA...27..272M, 2018ApJS..236....8L, 2019MNRAS.489..919K, 2020Natur.577..190M, 2020A&A...635A..61O, 2021arXiv210308410R}. Roughly a dozen FRBs have been localized to this precision and associated with a host galaxy with a spectroscopic distance. This sample of FRBs has been used to characterize the stellar environment of FRBs \citep{2021ApJ...908L..12T, 2021ApJ...917...75M}, study the FRB local magneto-ionic environment \citep{2018Natur.553..182M, 2021ApJ...908L..10H}, and measure the baryon density of the intergalactic medium \citep{2020Natur.581..391M}. 

Roughly 25 FRB sources are known to emit multiple bursts \citep[e.g.,][]{2019ApJ...885L..24C}, a few of which may exhibit periodic modulation to their burst rate \citep{2020Natur.582..351C, 2020MNRAS.495.3551R}. The discovery of repeating FRBs has had a large impact on the question of FRB origin, both because they demonstrate that some bursts are not cataclysmic and because repetition makes the sources easier to localize. Eight repeating FRBs have so far been localized precisely enough to be associated with a host galaxy \citep{2020Natur.577..190M, 2020ApJ...903..152H, 2021ATel14526....1L}. 

The first repeating FRB, known as 121102, was both the first to be localized and the first to be associated with a luminous persistent radio sources \citep[PRS;][]{2017Natur.541...58C}.
The PRS has a luminosity of L$_{\rm{1.4~GHz}} = 2\times10^{29}\,\rm{erg}\,\rm{s}^{-1}\,\rm{Hz}^{-1}$, comparable to low-luminosity AGN. The emission is isotropic, incoherent synchrotron radiation, which has been theoretically modeled to probe the FRB environment \citep{2016MNRAS.461.1498M, 2018ApJ...868L...4M, 2021MNRAS.501L..76K}. In this way, FRB 121102 motivated new, more detailed models for FRB origin through constraints on the characteristic age and energy density of the FRB environment. However, it remained the only example FRB/PRS association after the next dozen FRBs were localized, so its relevance to the overall FRB population was unclear. 

The recent discovery of FRB 20190520 (hereafter 190520) has changed this view dramatically. The source is similar to FRB 121102 in its burst activity, host galaxy properties, and association with a PRS \citep{niu_190520}. We now know that PRSs are an important part of the lives of some FRBs, but it is not clear how common they are. Given that FRBs occur with a high volumetric rate  \citep[comparable to that of core-collapse supernovae;][]{2018MNRAS.481.2320L, 2020ApJ...904...35P} and that PRS are luminous, it may be that PRS constitute a significant new class of extragalactic radio source.

A related open question is whether all FRBs are emitted by a single kind of source or that there are multiple sources of FRBs\footnote{We use the word ``source" to refer to a physical object formed in a particular way. By considering the formation channel in the definition of a source, one can distinguish between, say, a magnetar formed via a core collapse supernova and one formed via accretion-induced collapse.}. While it is currently not known if the origin of repeating FRBs and (apparent) non-repeaters are physically distinct, there is emerging evidence for differences in their burst properties. Repeating FRBs tend to be wider in duration and narrower in bandwidth than once-off events \citep{2019ApJ...885L..24C, Fonseca-2020, Pleunis-2021}. However, \citet{connor-2020} noted that viewing angle selection effects may explain effects like those observed. If PRS properties are causally connected to FRB properties, then it offers new ways to test the multiple-origin hypothesis.

Here, we consider the occurrence of PRS in FRBs, their prevalence in the local universe, and correlations between FRB and PRS properties. The goal of this analysis is to consider the FRB/PRS as a new class of radio source and discuss how to use them to test models of FRB origin. Section \ref{sec:pops} compiles measurements of the FRB/PRS population and suggests physically motivated definitions of subpopulations of FRBs. Section \ref{sec:occ} uses these definitions to demonstrate a preference for repeating FRBs to be associated with PRS. In Section \ref{sec:corr}, we discuss correlations between observed FRB and PRS properties. Section \ref{sec:src} discusses the volumetric density of the FRB population and Section \ref{sec:wild} discusses how PRS can be identified independently of FRBs.

\section{FRB and PRS Populations}
\label{sec:pops}

To begin, we compile measurements of FRB and PRS sources. Since the detection of FRB or PRS is sensitive to the quality of data, it is important have physically motivated definitions for the ``FRB'', ``PRS'', and ``repeating'' classes. PRS are most easily identified for localized FRBs, so we focus on that subset of all FRB sources. 

For the first ten years of study of FRBs, the practical definition of the source was a millisecond radio transient that was highly dispersed\footnote{Millisecond radio transients undergo a frequency-dependent time delay as they propagate through ionized gas. The frequency-dependent time delay is characterized by a dispersion measure (DM), which is equal to the integrated electron density $n_e$ along the line of sight (DM $\equiv \int_0^s n_e\, ds$).}. For FRBs with DM in excess of that expected from the Milky Way, their implied distance --- and luminosity --- was many orders of magnitude larger than that of Galactic millisecond transients like pulsars.

The discovery of a luminous radio burst from a magnetar in the Milky Way \citep[SGR 1935+2154;][]{2020Natur.587...59B, 2020Natur.587...54C} made it clear that FRBs reuqire an explicit luminosity definition. It has also become simpler to define the FRB class by its luminosity as more FRBs are associated with host galaxies. Therefore, we propose an FRB radio spectral energy threshold of $10^{29}$~erg Hz$^{-1}$, which includes all FRBs and excludes all other millisecond radio transients associated with well defined Galactic classes \citep[e.g., Crab giant pulses;][]{2004ApJ...612..375C,2021FrPhy..1624503L}. With this definition, a repeating FRB is defined as any source with multiple bursts with energy greater than $10^{29}$~erg Hz$^{-1}$.

A practical definition of a PRS should include the two, well characterized sources and exclude other classes of radio source. We suggest defining a PRS as an FRB associated radio source with a spectral luminosity\footnote{Here, we calculate a spectral luminosity for a flat spectral index: $L_\nu=4\pi \, D_L^2 \, S_\nu/(1+z)$, where $D_L$\, is the luminosity distance, $S_\nu$\, is the flux density, and $z$\, is the redshift.} L$_\nu>10^{29}\,\rm{erg}\,\rm{s}^{-1}\,\rm{Hz}^{-1}$ that is not attributed to star-formation activity in the host galaxy. Individual supernova remnants and star formation regions observed are far less luminous than this threshold \citep{2007ApJ...659..314P}. Radio emission related to active star-formation throughout the galaxy can be excluded through VLBI measurements or constraints on the star-formation rate \citep{Ravi+, Fong+}. The PRS luminosity limit also excludes all known typed supernovae  \citep[relativistic explosions such as SN Ic-BL are also excluded;][]{2021ApJ...908...75B}. However, this luminosity limit includes rare and extremely luminous radio transients \citep{2018ApJ...866L..22L, 2021arXiv210901752D}, typical active galactic nuclei \citep{2003MNRAS.345.1057M}, and broad regions allowed by supernova theory \citep{1998ApJ...499..810C, 2018ApJ...868L...4M, 2018MNRAS.474..573O}. 

In Table \ref{tab:frbprs}, we summarize the properties of FRBs with measurements or limits on PRS emission. We identified 24 FRBs that are either (1) localized to arcsecond precision or (2) have limits on the distance and associated radio flux density that place an upper limit on a PRS \citep{2020MNRAS.499.4716C, 2021arXiv210403991A}. The radio luminosity is recalculated from the flux density/distance limit assuming cosmological parameters defined in \citet{2016A&A...594A..13P}. The spectral properties of PRS are not well defined, so we identify a PRS by its spectral luminosity in units of erg s$^{-1}$~Hz$^{-1}$ and list the radio frequency of the measurement, $\nu_{\rm{PRS}}$.

\begin{table*}[htb]
\centering
 \begin{tabular}{llllllll}
 \hline
FRB & repeats & redshift/distance & DM & DM$_{\rm{host}}$ & L$_{\nu,PRS}$ & $\nu_{\rm{PRS}}$ & Refs \\
\hline
{\bf 121102} & yes & 0.1927 & 558.0 & 185.2 & 2.8e29 & 1.6 & C17 T17 L17 \\
{\bf 171020} & - & $<$0.08\tablenotemark{b} & 114.0 & -3.0 & $<$3.2e28 & 2.1 & M18 S18 \\
180301 & yes & 0.33 & 536.0 & 60.7 & $<$1.8e29 & 1.5 & Bh21a \\
{\bf 180309} & - & $<$0.32 & 263.0 & -87.4 & $<$4.6e28 & 3.0 & A21 \\
{\bf 180916B} & yes & 0.0337 & 349.0 & 74.6 & $<$4.9e26 & 1.7 & Mar20 \\
{\bf 180924} & - & 0.3214 & 361.42 & -12.7 & $<$5.7e28 & 6.5 & B19 Bh20a \\
{\bf 181030A} & yes & 0.00385 & 103.5 & 9.4 & $<$1.7e26\tablenotemark{a} & 3.0 & Bh21b \\
181112 & - & 0.4755 & 589.27 & 112.5 & $<$1.3e29 & 6.5 & H20 Bh20a \\
{\bf 190102} & - & 0.291 & 364.5 & 5.6 & $<$4.2e28 & 6.5 & H20 Bh20a \\
{\bf 190520} & yes & 0.241 & 1202.0 & 1097.4 & 3.0e29 & 3.0 & N21 \\
190523 & - & 0.66 & 760.8 & 129.1 & $<$4.3e30 & 3.0 & R19 \\
{\bf 190608} & - & 0.1178 & 339.5 & 171.2 & $<$3.8e27\tablenotemark{a} & 6.5 & Mac20 Bh20a \\
190611 & - & 0.378 & 321.4 & -163.8 & $<$2.9e30 & 0.9 & H20 RACS \\
190614D & - & $<$1.0 & 959.2 & -16.9 & $<$3.0e29 & 1.4 & L20 \\
190711 & yes & 0.522 & 587.4 & 22.4 & $<$5.6e30 & 0.9 & Ku21 Mac20 RACS \\
190714 & - & 0.2365 & 504.13 & 261.4 & $<$7.2e29 & 3.0 & H20 VLASS Bh19 \\
{\bf 191001} & - & 0.234 & 506.0 & 259.8 & $<$2.1e28\tablenotemark{a} & 5.5 & Bh20b \\
191108 & - & $<$0.52 & 588.1 & 112.0 & $<$2.6e30 & 1.4 & C20 \\
{\bf 191228} & - & 0.243 & 297.5 & 6.2 & $<$3.4e28 & 6.5 & Bh21a \\
{\bf 200120E} & yes & 3.6 Mpc & 87.0 & -3.7 & $<$3.1e23 & 1.5 & Ki21 \\
{\bf 200428} & - & 12.5 kpc & 332.0 & -0.0 & $<$2.0e23 & 1.4 & Ch20 B20 K18 \\
200430 & - & 0.16 & 380.1 & 194.0 & $<$3.2e29 & 3.0 & H20 VLASS Ku20 \\
{\bf 200906} & - & 0.3688 & 577.8 & 229.7 & $<$4.3e28 & 6.0 & Bh21a \\
{\bf 201124A} & yes & 0.098 & 420.0 & 232.6 & $<$2.8e28 & 1.4 & R21 F21 \\ \hline
\end{tabular}
\caption{Properties of localized FRBs with PRS measurements. FRB names in bold have PRS detections or meaningful upper limits. References: C17 \citep{2017Natur.541...58C}, L17 \citep{2017ApJ...850...76L}, R16 \citep{2016Sci...354.1249R}, T17 \citep{2017ApJ...834L...7T}, M18 \citep{2018ApJ...867L..10M}, S18 \citep{2018Natur.562..386S}, A21 \citep{2021arXiv210403991A}, M20 \citep{2020Natur.577..190M}, B19 \citep{2019Sci...365..565B}, R19 \citep{2019Natur.572..352R}, M20 \citep{2020Natur.581..391M}, C20 \citep{2020MNRAS.499.4716C}, L20 \citep{2020ApJ...899..161L}, CH20 \citep{2020Natur.587...54C}, B20 \citep{2020Natur.587...59B}, Bh20a \citep{2020ApJ...895L..37B}, Bh20b \citep{2020ApJ...901L..20B}, H20 \citep{2020ApJ...903..152H}, K19 \citep{2019ApJ...887L..30K}, K18 \citep{2018ApJ...852...54K}, Ku21 \citep{2021MNRAS.500.2525K}, Ki21 \citep{2021arXiv210511445K}, R21 \citep{Ravi+}, F21 \citep{Fong+}, Bh21a \citep{2021arXiv210801282B}, Bh21b \citep{2021arXiv210812122B}, Bh19 \citep{2019ATel12940....1B}, Ku20 \citep{2020ATel13694....1K}, RACS \citep{10.1017/pasa.2020.41}, VLASS \citep{2020PASP..132c5001L}}
\tablenotetext{a}{PRS limit set by radio emission attributed to star formation activity in host galaxy.}
\tablenotetext{b}{We assume a max distance of $z=0.08$ for FRB 171020, which is more conservative than presented in \citet{2018ApJ...867L..10M}.}
\label{tab:frbprs}
\end{table*}

Four FRBs (190611, 190711, 190714, 200430) are well localized, but have no published limit on the flux density of radio counterparts. In the northern sky, the positions of FRBs 190714 and 200430 were observed by the VLA Sky Survey \citep{2020PASP..132c5001L} and in the southern sky, the positions of FRBs 190611 and 190711 were observed by the ASKAP RACS survey \citep{10.1017/pasa.2020.41}. No counterparts were found for any of these FRBs, which sets a $3\sigma$~ limit of 0.5 mJy (at 3 GHz) for 190714 and 200430  and 0.75 mJy (at 900 MHz) for 190611 and 190711. FRB 200428 (a.k.a. SGR 1935+2154) is in the Milky Way, but we consider it with an extragalactic perspective. We conservatively associate it with persistent radio emission from a $\sim$20~pc supernova remnant  \citep{2018ApJ...852...54K} \footnote{Limits on more compact radio emission are five orders of magnitude lower \citep{2020ATel13693....1R}.} and treat it as a non repeater with no PRS, according to our definitions. Four FRBs have persistent radio emission associated with star formation: 191001 \citep{2020ApJ...901L..20B}, 190608 \citep{2020ApJ...895L..37B}, 201124A \citep{Ravi+, Fong+}, and 181030A \citep{2021arXiv210812122B}. In these cases, the upper limit on PRS emission associated with the FRB is less than the PRS luminosity threshold, so the presence of a PRS is excluded.

For each source, we summarize the observed DM and an estimate of the DM that can be attributed to its host galaxy, including the galaxy halo, ISM, and the FRB local environment. The DM contribution can be expressed as a sum of physically distinct components:
\begin{equation}
\rm{DM} = \rm{DM}_{\rm{MW}} + \rm{DM}_{\rm{MW,halo}} + \rm{DM}_{\rm{IGM}} + DM_{\rm{host}}/(1+z),
\end{equation}
\noindent with DM$_{\rm{host}}$\ defined in the rest frame. The Milky Way halo DM has both theoretical and data-driven estimates that are consistent with DM$_{\rm{MW,halo}}=50\,\rm{pc}\,\rm{cm}^{-3}$ \citep{2020MNRAS.496L.106K, 2020ApJ...895L..49P, 2021MNRAS.500..655D}. We calculate DM$_{\rm{MW}}$\ using the NE2001 Milky Way electron density model \citep{2002astro.ph..7156C}. DM$_{\rm{IGM}}$\ is calculated from a model of the average IGM \citep{2016A&A...594A..13P, 2019zndo...3403651P}.

\begin{figure}[htb]
    \centering
    \includegraphics[width=\columnwidth]{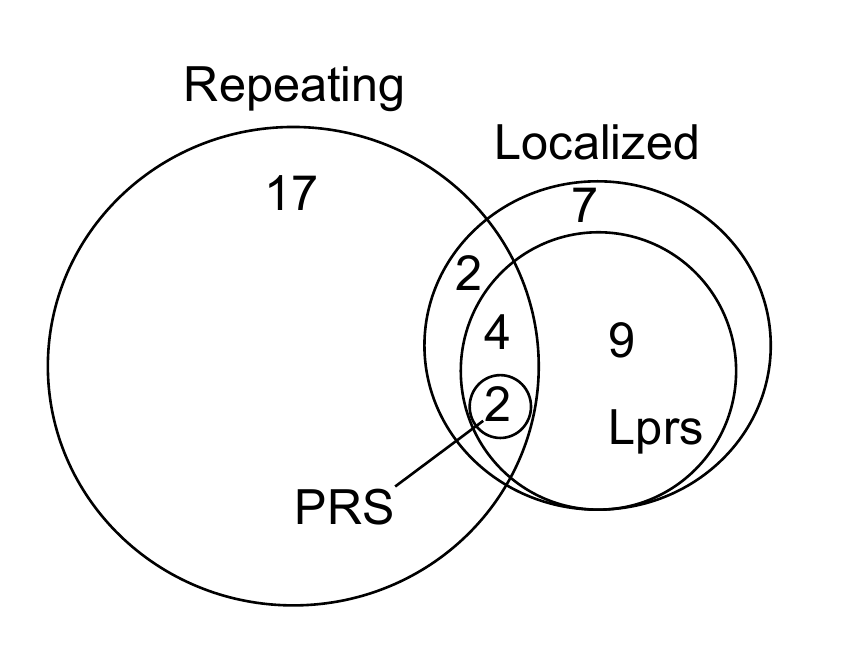}
    \caption{A Venn diagram showing how FRBs can be assigned to subgroups. The ``repeating'' circle includes 21 repeating FRBs listed in the Transient Name Server (\url{http://wis-tns.org}). The ``localized'' circle includes all sources shown in Table \ref{tab:frbprs}. The ``Lprs'' circle includes the subset of localized FRBs with detections or luminosity limits L$_{\rm{r,PRS}}<=10^{29}\,\rm{erg}\,\rm{s}^{-1}\,\rm{Hz}^{-1}$; this sample is also shown in Figure \ref{fig:prsfrac}. Finally, the ``PRS'' circle shows the two localized FRBs with PRSs.}
    \label{fig:frbvenn}
\end{figure}

Given these classes, we show the FRB subgroups as a Venn diagram in Figure \ref{fig:frbvenn}. Although Table \ref{tab:frbprs} includes 24 FRBs, a subset of 15 (shown with names in {\bf bold}) have radio limits sensitive enough to detect a PRS. FRB 181112 is close to the limit, but formally counted as unconstrained. FRBs 171020 and 180309 are not localized, but have limits on bright radio counterparts that excludes association with a PRS. We can exclude the association of the whole sample of 24 FRBs with a PRS L$_{\rm{r,PRS}}>10^{31}\,\rm{erg}\,\rm{s}^{-1}\,\rm{Hz}^{-1}$. While a threshold of L$_{\rm{r,PRS}}>10^{28}\,\rm{erg}\,\rm{s}^{-1}\,\rm{Hz}^{-1}$ would measure or constrain a PRS for five repeating FRBs and three non-repeating FRB. The sources identified as PRS/non-PRS do not change when scaling luminosity limits to a frequency of 3 GHz with a typical synchrotron spectral index of $-0.7$. 

\section{PRS Occurrence in FRBs}
\label{sec:occ}

With the classification of localized FRBs into the observational categories of repeater/non-repeater and PRS/no-PRS, we can calculate the fraction of sources in these classes. Specifically, we want to know what fraction of repeaters, $f_{r}$, has a detectable persistent source relative to the fraction for non-repeaters, $f_{nr}$. Current data allow us to estimate the true $f_{r}$ and $f_{nr}$ values of the underlying population. We model the question with a binomial distribution. This is analogous to a biased coin flip, where one can ask ``If I flip a coin $n$ times and it comes up heads $k$ times, what are the allowed values for the probability of heads on this coin?''. Similarly, if we observe $n_{\rm nr}$ non-repeating FRBs and none has a PRS, we want to know the maximum allowed value of $f_{\rm nr}$. For repeating FRBs, we can calculate the allowed values of $f_{\rm r}$, given $k_{\rm r}$ out of $n_{\rm r}$ repeaters has a coincident PRS. The following two equations give the probability distributions for the PRS fraction $f$ for repeaters and non-repeaters,

\begin{equation}
p(f_{\rm r} | n_{\rm r},\,k_{\rm r}) = {n_{\rm r} \choose k_{\rm r}}  f_{\rm r}^{k_{\rm r}} (1 - f_{\rm r})^{n_{\rm r}-k_{\rm r}}
\end{equation}

\begin{equation}
p(f_{\rm nr} | n_{\rm nr},\,k_{\rm nr}\!=\!0) = {n_{\rm nr} \choose 0} \left (1 - f_{\rm nr} \right )^{n_{\rm nr}}
\end{equation}

\begin{figure}[htb]
    \centering
    \includegraphics[width=\columnwidth]{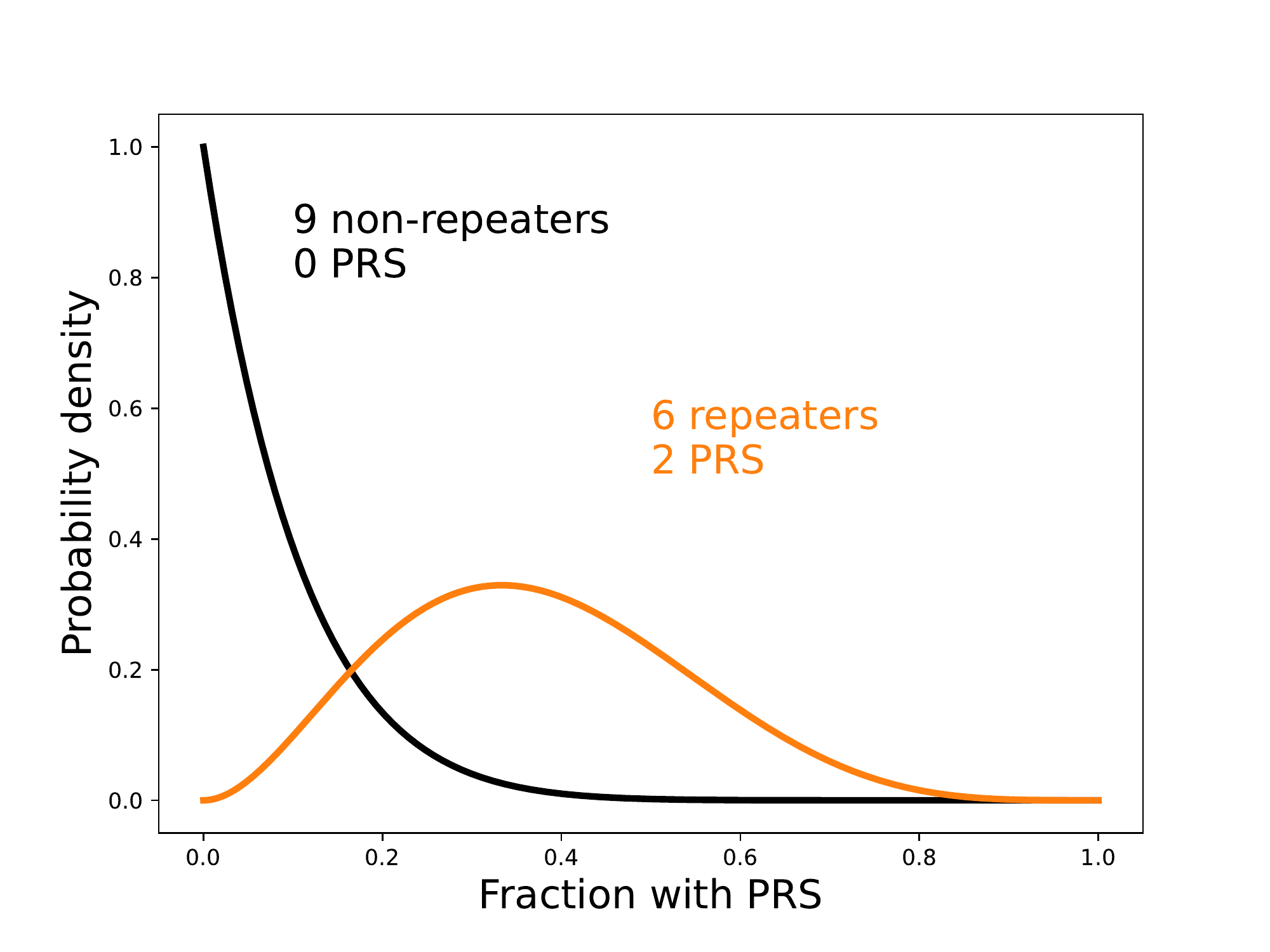}
    \caption{A comparison of PRS occurrence in the nine localized FRBs with deep radio imaging sensitive to PRS luminosities of $10^{29}\,\rm{erg}\,\rm{s}^{-1}\,\rm{Hz}^{-1}$. Curves show the binomial distribution for the repeating (orange) and non-repeating (black) groups.
    \label{fig:prsfrac}}
\end{figure}

Currently, there are 9 localized non-repeaters sensitive to a PRS. Given none has been detected, we place a $90\%$ upper limit of $f_{\rm nr}<$0.23. There are two localized repeaters with PRS out of 6 searched, which gives a $90\%$ confidence region of $0.15<f_{\rm r}<0.73$. The probability distributions for the two observational classes are shown in Fig.~\ref{fig:prsfrac}. With current data, there is no strong evidence that PRS are preferentially found coincident with 
repeating FRBs. As a simple estimate, consider that the probability of two PRS showing
up in a 6-FRB subset out of 15 total is given by $\frac{6}{15}\times\frac{5}{14}\approx0.14$. If 
we also sum over cases as extreme or more extreme than this one (i.e. $\geq2$ out of 6), we find a probability closer 
to 0.20. 
But this estimate could be significantly improved with radio imaging of the 8 localized FRBs that have poor constraints on L$_{\rm{r,PRS}}$.





We emphasize that without further considerations, this question is strictly observational and not necessarily physical. For example, if repeating FRBs were systematically more nearby or had more sensitive images, then their PRS would be more easily detectable than non-repeating FRBs. In the following section, we show that the PRS measurements are not strongly biased in this way.

Considering both repeating and non-repeating FRBs, there are two of 15 with PRS counterparts, which gives PRS fraction of $0.06<f_{\rm{all}}<0.36$ (90\% confidence). This fraction is subject to an additional bias, which is the relative likelihood of localizing repeating FRBs relative to non-repeating FRBs. 
Assuming that all FRBs repeat at some level, then the chance of finding a FRB is higher for FRBs that are brighter or more active. No additional bias is introduced for identifying a PRS, beyond that of FRB localization, so the PRS fraction is appropriate for the sample of detected FRBs.

\section{What FRB Properties Predict the Presence of a PRS?}
\label{sec:corr}

\begin{figure*}[htb]
    \centering
    \includegraphics[width=\textwidth]{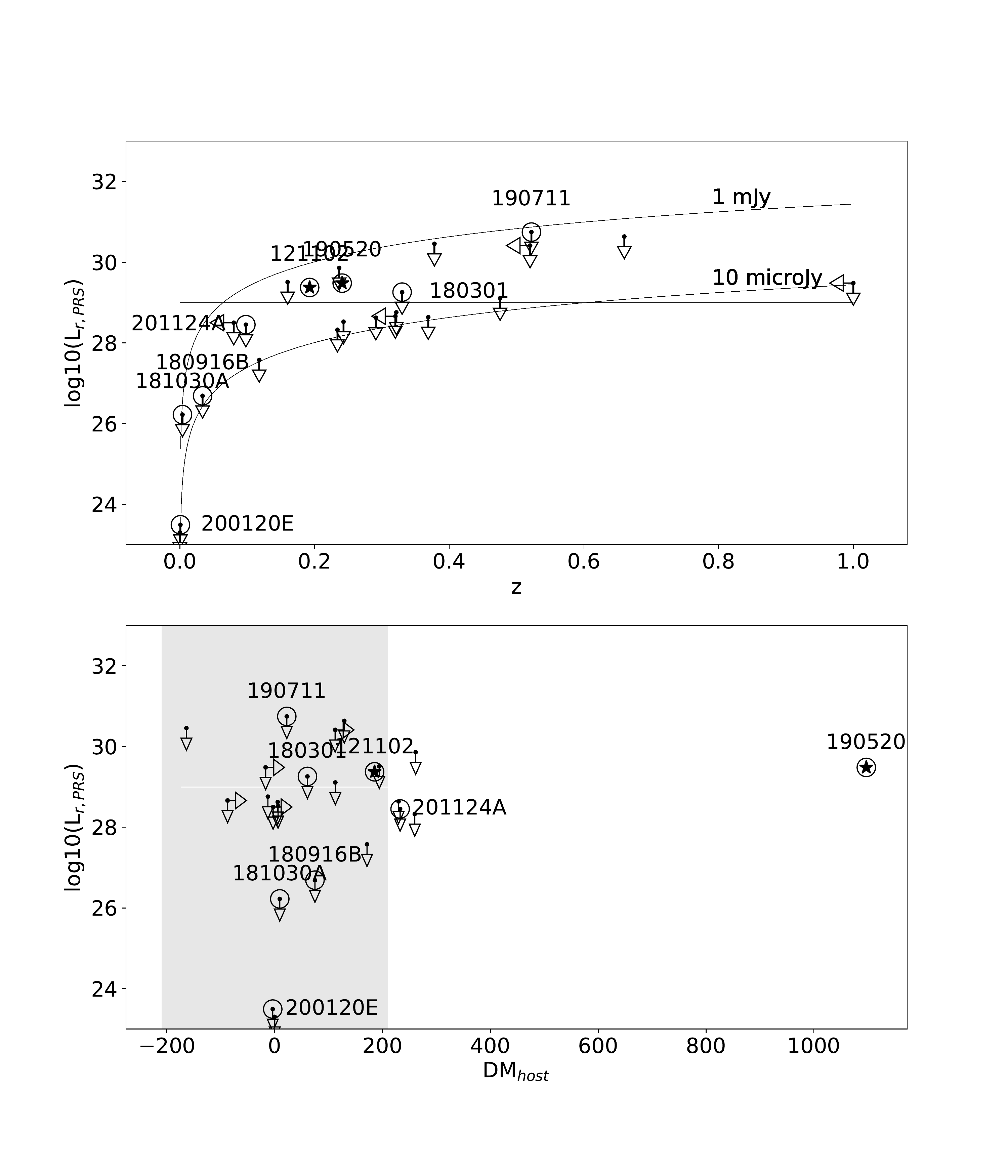}
    \caption{(Top:) PRS luminosity measurements or limits as a function of host galaxy redshift. Repeating FRBs are circled and labeled by name; PRS detections shown as a star. The solid horizontal line shows the defined luminosity for a PRS, L$_{\rm{PRS}}=10^{29}\,\rm{erg}\,\rm{s}^{-1}\,\rm{Hz}{-1}$. The two curved lines show the luminosity limit for flux density limits of 1 mJy and 10 microJy. (Bottom:) PRS luminosity versus DM$_{\rm{host}}$\ (rest frame) for the same FRB sample. FRB 200430 (SGR 1935+2154) is assigned a DM$_{\rm{host}}$\ of 0, which partially overlaps with FRB 200120E. The shaded region shows $|$DM$_{\rm{host}}|<210\,\rm{pc}\,\rm{cm}^{-3}$, which is a typical contribution of the ISM of a host galaxy.
    \label{fig:lprs}}
\end{figure*}

Here we consider FRB properties that might explain their association with a PRS. Figure \ref{fig:lprs} shows correlations between L$_{\rm{r,PRS}}$, DM$_{\rm{host}}$, and repetition\footnote{An assumption of the this analysis is that the burst repetition and PRS luminosity are independent and that the PRS is not \emph{exactly equal to} the integrated burst emission. \citet{2019ApJ...877L..19G} considered whether the FRB 121102 bursts can produce its PRS. They found that 700 bursts/ms are required to produce the PRS luminosity, which is not consistent with the observed burst energy distribution.}. DM$_{\rm{host}}$ itself can be attributed to three distinct components: host galaxy halo, ISM, and the local FRB environment. The first two were considered in \citet{2020Natur.581..391M}, which used localized FRBs to estimate the baryon density of the IGM. They defined a probability distribution for DM$_{\rm{host}}$ as:
\begin{eqnarray}
    p_{\text{host }}(\mathrm{DM}_{\text{host}}\mid \mu,\sigma_{\text {host}}) = && \frac{1}{(2 \pi)^{1 / 2} \mathrm{DM} \sigma_{\text{host}}} \nonumber
    \\
    && 
    \exp \left[-\frac{(\log \mathrm{DM}-\mu)^{2}}{2 \sigma_{\text {host }}^{2}}\right]
\end{eqnarray}
\noindent where $\mu$ and $\sigma_{\rm{host}}$ represent the mean and standard deviation of host galaxy DM.

This functional form was designed to estimate the host galaxy halo and ISM contributions, not local DM. Furthermore, the FRB sample definition excluded FRB 121102 due to its anomalously large (presumably local) DM. This model finds a rest-frame halo/ISM DM contribution with mean 68 pc cm$^{-3}$\ and width parameter of 0.88, which predicts DM$_{\rm{host}}<210$ (90\% confidence) for the typical FRB host galaxy. This is consistent with detailed DM modeling of foreground galaxies toward FRB 180924 and 190608, which lie along relatively under- and over-dense lines of sight, respectively \citep{2021arXiv210809881S, 2020arXiv200513157S}.

The bottom panel of Figure \ref{fig:lprs} shows how the measured DM$_{\rm{host}}$ compares to that expected from the typical FRB host. FRBs 121102 and 190520 stand out for having high PRS luminosities, large DM$_{\rm{host}}$, and high burst rates. In contrast, FRBs 180916B and 201124A actively repeat, but have limits on PRS emission. Clearly, repetition rate is not sufficient to predict the presence of a PRS. However, FRBs with both a large repetition rate and DM$_{\rm{host}}$ may be more likely to have a PRS. The large DM$_{\rm{host}}$ of FRB 201124A may seem to contradict this point, but we note that the host galaxy is more massive and has more star formation than the typical FRB host.
To quantify this further, we normalized DM$_\mathrm{host}$ by host galaxy stellar mass and star formation rate \citep{2020ApJ...903..152H, niu_190520}. Normalizing by stellar mass shows that FRB\,121102 and 190520 are outliers by an order of magnitude. The distinction is not as clear for DM$_\mathrm{host}$ normalized by SFR, with FRB\,12112, 190520, 180916 and 190523 showing relatively high values.

If we can associate large DM$_{\rm{host}}$\ to sources with large PRS luminosity, it would have tremendous diagnostic power. The PRS is emitted by a relativistic plasma, while DM is caused by a physically distinct cold plasma. Some FRB source models predict local counterparts, such as HII regions, SNRs, and PWNe, that can contribute significant local DM and relativistic plasma \citep[$>$100 pc cm$^{-3}$;][]{2016MNRAS.458L..19C, 2017ApJ...847...22Y}. Young magnetar and fast pulsar models have been used for detailed calculations of ionization and radiation calculations that predict significant contribution to FRB DM and a luminous PRS \citep{2017ApJ...839L...3K, 2018ApJ...868L...4M}. On a larger scale, calculations of SN ionized ejecta and stellar winds predict DM as high as 100--1000 pc cm$^{-3}$ and RM of 10$^4$--10$^6$ rad m$^{-2}$ after 10-100 years \citep{2018ApJ...861..150P}. Milky Way pulsars in SNRs have an excess DM \citep{2020A&A...634A.105S} that is consistent with predictions of much larger excess in the younger scenarios predicted for FRBs. Alternatively, compact binary mergers have a distinctive scale for DM and RM evolution \citep{2021ApJ...907..111Z}. 


If DM$_{\rm{host}}$\ is correlated with PRS luminosity and repetition rate, then we can use this to guide follow-up observing. Table \ref{tab:frbprs} and Figure \ref{fig:lprs} highlight FRB sources with large DM$_{\rm{host}}$ ($>150\,\rm{pc}\,\rm{cm}^3$) and weak constraints on PRS emission: FRBs 190714 and 200430. Deep radio imaging and/or and FRB search of these sources may be more likely to find new bursts and/or PRS emission.

Finally, we note that the top panel of Figure \ref{fig:lprs} demonstrates how PRS limits scale with distance and helps demonstrate potential bias between the repeater/non-repeater populations. The PRS luminosity limit tends to be derived from shallow all-sky surveys or deep follow-up imaging (characteristic sensitivities are shown as lines in the Figure). All-sky surveys have a completeness limit of roughly 1 mJy, which is sensitive to the PRS luminosity to a distance of $z=0.065$\ (L$_d=300$~Mpc), while a 1-hour, 1.4-GHz VLA observation has a sensitivity of 10 microJy and is sensitive to the PRS luminosity to $z=0.6$\ (L$_d=3.6$~Gpc). Most FRBs with meaningful PRS limits would also limit PRS emission out to $z\approx0.4$, a distance that includes a similar number of repeating and non-repeating FRBs. This supports the argument that PRS emission from repeaters and non-repeaters is equally well constrained and that the preference for PRS emission in repeating FRBs is not strongly biased by distance (Figure \ref{fig:prsfrac}).

\section{FRB Source Density}
\label{sec:src}

With a working definition of a PRS, we can consider how they contribute to radio source counts. First, we need to estimate the FRB source density from the volumetric rate and repetition statistics. This estimate uses data to bound the repetition rate, which was a parameter in previous estimates \citep{2017ApJ...843...84N}.

The volumetric density of FRB emitting sources can be defined as,

\begin{equation}
  \mathcal{N}_{\rm src} = \frac{\Phi_{\rm z}}{\left<\mathcal{R}_{\rm src}\right> f_{\rm{b}}}
\end{equation}

\noindent where $\Phi_{\rm z}$ is the FRB volumetric rate at redshift z (in units of Gpc$^{-3}$\ yr$^{-1}$), $\left<\mathcal{R}_{\rm src}\right>$ is the average burst rate per source, and $f_{\rm{b}}$\ is the beaming fraction of FRB emission (fraction of sources with beams pointed toward us). The volumetric rate can be estimated from the detection rate of all FRBs combined with knowledge of the redshift distribution of FRBs. The estimate of this rate has been approached with a variety of techniques and data sets and are generally consistent \citep{2019MNRAS.487.5753C, 2019ApJ...883...40L, 2020MNRAS.494..665L, 2021MNRAS.501.5319A, 2021A&A...647A..30G}. We use the rate of \citet{2021arXiv210107998J}, which is  $\Phi_{\rm{z=0}}=9^{+2}_{-4}\times10^4$\ Gpc$^{-3}$\ yr$^{-1}$ with burst energy greater than $10^{39}$\,erg in the local universe.
 
The average burst rate per source is more difficult to entertain, because we are forced to calculate an expected value over a broad distribution of repeat rates, with significant weight at $\mathcal{R}_{\rm src}=0$ if some FRBs are true non-repeaters. We simplify the problem by considering the limiting cases that either (1) all FRBs repeat or that (2) there is a mix of repeaters and true non-repeaters. We use the observed CHIME/FRB burst sample under these two cases to bound the estimated average repetition rate of FRBs. This estimate is based on the 20 publicly available CHIME/FRB repeating sources\footnote{\url{https://www.chime-frb.ca/repeaters}} and the 474
non-repeaters in the CHIME/FRB catalog.

As a lower limit on $\left<\mathcal{R}_{\rm src}\right>$, assume that all sources only detected once at CHIME/FRB are true non-repeaters with rate zero. Then assume there are no values of $\mathcal{R}_{\rm src}$ higher than the most active CHIME repeater. By a simple weighted average, this gives:

\begin{equation}
    \left<\mathcal{R}_{\rm src}\right>_{\rm low} = \frac{1}{N_{\rm R} + N_{\rm NR}}
    \left (N_{\rm R}\left<\mathcal{R}_{\rm R}\right> + N_{\rm NR}\left<\mathcal{R}_{\rm NR}\right> \right ),
\end{equation}

\noindent where $N_{\rm R}$ and $N_{\rm NR}$ are the numbers of repeaters and non-repeaters in the CHIME/FRB sample. 
The available CHIME/FRB repeaters have a mean 
repeat rate of roughly 580 bursts per year, assuming each has been 
observed since August 2018 for 10 minutes per source per day. Using 
474 non-repeaters and 20 repeaters, we find $\left<\mathcal{R}_{\rm src}\right>_{\rm low}\approx$\,25\,yr$^{-1}$. 


The upper limit on the average rate (and lower bound on the density of sources) comes from assuming that all FRBs detected by CHIME/FRB are repeaters, but the survey has not been on sky long enough to detect some of them more than once. We assume that all FRBs obey a power-law distribution in repetition rate with a probability density function, $n(\mathcal{R})\propto\mathcal{R}^{-\alpha}$. We determine $\alpha$ empirically from the CHIME/FRB repeaters to be $\approx$\,1.5, using the maximum-likelihood estimator 
from \citet{crawford-1970}. The expected value in this case is

\begin{equation}
    \left<\mathcal{R}_{\rm src}\right>_{\rm up} =  \int^{\mathcal{R}_{max}}_{\mathcal{R}_{min}} \mathcal{R}\,n(\mathcal{R}) \,\, \mathrm{d}\mathcal{R}.
\end{equation}

\noindent Replacing $n(\mathcal{R})$ with an integration constant times the power-law, $C\, \mathcal{R}^{-\alpha}$, we get

\begin{equation}
    \left<\mathcal{R}_{\rm src}\right>_{\rm up} =  \frac{C}{2-\alpha}\,\left ( \mathcal{R}^{2-\alpha}_{max} - \mathcal{R}^{2-\alpha}_{min} \right ).
    \label{eq:int}
\end{equation}

We take $\mathcal{R}_{max}$ to be the highest rate of any CHIME FRB, and $\mathcal{R}_{min}$ to be the reciprocal of the total exposure per source since CHIME/FRB first light, $1/T_{\rm tot}$. To reiterate, in this scenario we are assuming that all CHIME sources are repeaters, and the FRBs that have only been detected once simply have low activity with $\mathcal{R}\lesssim2/T_{\rm tot}$. Plugging these values into Eq.~\ref{eq:int} gives $\left<\mathcal{R}_{\rm src}\right>_{\rm up}=440$\,yr$^{-1}$. 

Given the bounds on repetition and the volumetric FRB rate, we can calculate the density of FRB-emitting sources. Using the 2$\sigma$\ range on the volumetric rate, we bound the source density, $2.2\times\,10^2 f_{\rm{b,0.1}}^{-1}$\ Gpc$^{-3}$ $<\mathcal{N}_{\rm{src}}<5.2\times\,10^{4} f_{\rm{b,0.1}}^{-1}$\ Gpc$^{-3}$, where $f_{\rm{b,0.1}}= f_{\rm{b}}/0.1$ is a parameterized beaming fraction. The value of $f_{\rm{b}}$ can vary widely with emission mechanism, with a characteristic value of 1/10 for pulsars \citep{1998MNRAS.298..625T, 2010ApJ...715..230O} and potentially much smaller values for coherent emission mechanisms \citep{2016MNRAS.457..232C,  2017MNRAS.467L..96K}.

This constraint is built on assumptions of burst energy and repetition. This FRB volumetric rate is based on an energy limit ($E>10^{39}$\ erg) that corresponds to a burst fluence of 15 Jy~ms in 256~MHz at $z=0.1$. This distance is roughly equal to that which defines a complete volume for the sample of pre-commissioning FRBs seen by CHIME/FRB, as discussed in \citet{2019NatAs...3..928R}. With a lower flux density threshold, a more powerful survey could go further down the FRB luminosity function and perhaps have access to more sources in a given volume. The repetition statistics are derived from CHIME/FRB sources, which may differ from that seen by other telescopes (e.g., based on pulse width, DM/scattering distribution, emission frequency).

\section{PRS in the Wild}
\label{sec:wild}

PRS are as luminous as AGN, the most common class of extragalactic radio source \citep{2019ApJ...872..148C}. Given that AGN are detectable at great distance, it is reasonable to wonder whether FRB sources may be detectable via their PRS counterparts. Here, we use the bound on FRB source density from \S \ref{sec:src} to estimate the density of PRS, their contribution to radio source catalogs, and prospects for identifying them independent of their FRB emission.

For a FRB source density of $\mathcal{N}_{src}$, we expect a PRS density of $\mathcal{N}_{\rm{PRS}}=f_{\rm{all}} \mathcal{N}_{src}$. In Section \ref{sec:occ}, we estimated the PRS occurrence to be between 0.06--0.36, which we paramterize to a value of $f_{\rm{all}}=0.2$. Thus, we find $\mathcal{N}_{\rm{PRS}} \approx 50 - 10000\, f_{\rm{b}, 0.1}^{-1}\, f_{\rm{all},0.2}$\ Gpc$^{-3}$. Table \ref{tab:rates} summarizes the bounds on FRB and PRS densities and rates.

At this density, radio source catalogs and new sky surveys should detect a significant number of PRS. The FIRST, VLASS, and RACS radio surveys are complete to a flux density of roughly 1 mJy, which detects all PRS out to a distance of $z=0.065$\ (luminosity distance of 300 Mpc). For an all-sky survey (typically seeing 3$\pi$ steradians), these surveys are sensitive to these sources in a volume of 0.08 Gpc$^3$, which includes $4 - 830\, f_{\rm{b}, 0.1}^{-1}\, f_{\rm{all},0.2}$\ PRS. This likely underestimates the number of detectable sources, because it assumes a fixed PRS luminosity of $10^{29}\,\rm{erg}\,\rm{s}^{-1}\,\rm{Hz}^{-1}$. The two known PRS have luminosities three times larger than that and would be detectable over a larger volume.

By number, PRSs are a small fraction ($10^{-5}$\ -- $10^{-3}$) of a typical radio catalog \citep[e.g., VLASS epoch 1 catalog has 1.7$\times10^6$\ sources;][]{2021arXiv210211753G}. \citet{2019ApJ...872..148C} use radio and infrared emission to classify extragalactic radio sources in the local universe. For luminosities greater than $10^{29}\,\rm{erg}\,\rm{s}^{-1}\,\rm{Hz}^{-1}$\, the density of star-forming galaxies and AGN is $1.4\times10^5$\ and $1.0\times10^6\,\rm{Gpc}^{-3}$, respectively. So in the local universe, PRS potentially amount to as much 1\% and 7\% of the radio-luminous AGN and star-forming galaxy populations, respectively.

\section{Discussion and Conclusions}

A remarkable fact about FRBs is that their volumetric rate is roughly equal to that of core-collapse supernovae, $\Phi_{\rm{ccsn}}=10^{+5}_{-3.5}\times10^4$\ Gpc$^{-3}$\ yr$^{-1}$ \citep[at $z=0$;]{2015A&A...584A..62C, 2020ApJ...904...35P}. The rate scales with energy as $R(>E_\nu)\propto E_\nu^{-1}$, so low-energy FRBs are more abundant \citep{2021arXiv210704059L}. Plausible progenitor channels are too rare to explain the FRB volumetric rate \citep{2019NatAs...3..928R} and sources at low energy must either emitted by more abundant sources or have a higher repetition rate \citep{2021arXiv210812122B}.

The most detailed physical models for repeating FRBs are based upon magnetars \citep{2017ApJ...839L...3K, 2018ApJ...868L...4M}. The magnetars that produce flaring events as soft-gamma repeaters (SGR) have been characterized well. SGR flares occur at a rate of $3.8^{+4.0}_{-3.1}\times10^5\ \rm{Gpc}^{-3}\ \rm{yr}^{-1}$ above an energy of $4\times10^{44}$ erg \citep{2021ApJ...907L..28B}, which is comparable to the FRB rate above $10^{39}$ erg \citep{2021arXiv210107998J}. The ratio of energy in these two bands, referred to as $\eta$, is $4\times10^5$, which is consistent with some theoretical models and observational constraints \citep{2020ApJ...897..146C, 2021NatAs...5..401T}.


In estimating the mean FRB repetition rate, we have considered two scenarios: (1) all FRBs repeat with a rate drawn from a powerlaw distribution and (2) some FRBs repeat and some are cataclysmic. Analysis of FRB spectra and temporal widths have also been used to argue that repeating and non-repeating FRBs are distinct classes \citep{2019ApJ...885L..24C, Fonseca-2020, Pleunis-2021}. Here, we identify PRS emission as a potential new signature of a distinct subpopulation of FRBs. If true, PRS is likely to be more useful in defining classes, since it can more easily be connected to simple physical parameters \citep[e.g., energy input, age;][]{2017ApJ...839L...3K, 2017ApJ...841...14M}.

As more sources are localized and more FRB local environments are studied, other relationships may become apparent (e.g. Faraday rotation measure, scattering, stellar environment).
The stellar environment of the FRB is also easier to interpret physically \citep[e.g., mean age of progenitors;][]{2021ApJ...917...75M}. Generally, the distribution of FRB host galaxy properties suggest that FRBs trace star formation rate \citep{2021ApJ...907L..31B}. However, if FRBs with PRS tend to form in small, star-forming galaxies, it may favor the idea that they are formed by a process that differs from that of non-repeating FRBs.

In this work, we bound the FRB source density in order to compare it to other classes of object. Table \ref{tab:rates} summarizes bounds on rates and densities for both FRB sources and PRS. The FRB source density estimate is consistent with that of \citet{2016MNRAS.461L.122L}, which assumed that all FRB sources are emitted by neutron stars and repeat with some universal energy distribution.

\begin{table}[tb]
\centering
 \begin{tabular}{lll}
 \hline
  Parameter & Symbol & Range (90\% CI) \\ \hline
  Repetition rate & $\left<\mathcal{R}_{\rm src}\right>$ & 25 -- 440 yr$^{-1}$  \\
  FRB source density &  $\mathcal{N}_{src}$ & 220 -- 52000 $f_{\rm{b}, 0.1}^{-1} \rm{Gpc}^{-3}$ \\
  PRS fraction & $f_{\rm{all}}$ & 0.06 -- 0.36 \\
  PRS density & $\mathcal{N}_{\rm{PRS}}$ & 50 -- 10000 $f_{\rm{b}, 0.1}^{-1}\, f_{\rm{all},0.2} \rm{Gpc}^{-3}$  \\
 \end{tabular}
 \caption{FRB and PRS Rates and Densities. Estimate is appropriate for sources that emit FRBs with an energy $>10^{39}$\ erg at $z=0$, as calculated in \citet{2021arXiv210107998J}.}
 \label{tab:rates}
\end{table}
 
The consistency of burst rates with predictions from magnetar models belies the fact that magnetars are far more common than FRB sources. Magnetars constitute a significant fraction ($\sim0.4$) of the neutron star population and are born at a rate of roughly $10^{-2}\ \rm{yr}^{-1}$ in the Milky Way \citep{2019MNRAS.487.1426B}. Given that most of their energy is stored in magnetic field that decays on a timescale of $10^4$ yr, roughly 100 magnetars are present in a Milky Way-like galaxy at any time. The density of Milky Way-like galaxies is $\Phi_{\rm{MW}} \approx 10^{7}\ \rm{Gpc}^{-3}$ \citep{2003ApJ...592..819B}, so the density of magnetars is roughly $10^{9}\ \rm{Gpc}^{-3}$. This is a factor of $10^{6}$ larger than the FRB source density \citep[consistent with independent estimates;][]{2021arXiv210704059L}. If magnetars do emit FRBs, then they must occur in a tiny subset of all magnetars, such as the youngest or most magnetic sources.


We have also used the prevalence of PRS in FRBs to estimate their contribution to the persistent radio source population. 
By definition, PRS are associated with FRBs, so this provides a complementary way to test models for FRB origin. Interestingly, PRS look remarkably similar to compact, low-luminosity AGN. As discussed above, the volumetric density of galaxies (and their supermassive black holes) is many orders of magnitude larger than the PRS density, but subclasses do occur at a comparable level \citep[e.g, AGN identified via their broad-line emission;][]{2007ApJ...667..131G}.

PRS are distinctive for being compact, radio luminous, and associated with star-formation \citep{2021ApJ...908L..12T, 2021ApJ...917...75M}. However, it is important to note that the non-nuclear location of the two known PRS does not preclude their association with AGN. \citet{2020ApJ...888...36R} identified dozens of luminous (some non-nuclear) radio sources in dwarf galaxies that are consistent with AGN, so-called ``wandering black holes''. 
This search was complete to VLA/FIRST radio sources in dwarf galaxies within 225 Mpc and identified five with luminosities consistent with a PRS. 
Given the footprint of the surveys, we estimate a volumetric density of $4\times10^2\ \rm{Gpc}^{-3}\ \rm{yr}^{-1}$. Therefore, the density of PRS is consistent with that of candidate AGN in dwarf galaxies. \citet{2020ApJ...895...98E} also detailed the phenomenological similarity of the PRS associated with FRB 121102 to the sample of \citet{2020ApJ...888...36R}. 

 An FRB maximalist may be tempted to categorize ``wandering black holes'' as PRS. However, there is sigificant theoretical and observational motivation for the existence of off-nuclear AGN \citep{2020ARA&A..58..257G, 2021ApJ...913..102W}. In many cases, the luminous radio emission can be directly attributed to accretion onto a black hole \citep{2020ApJ...898L..30M, 2021ApJ...910....5M}. Given the uncertain nature of PRS, it is also possible that off-nuclear AGN produce PRS and FRBs \citep[e.g.,][]{2020Natur.587...45Z}!

The ambiguity between AGN and PRS argues for caution when classifying based on radio data alone. Nuclear radio sources are likely to be AGN and off-nuclear radio sources may be a PRS, but gas dynamics and ionization are required to show that definitively. Given the similar densities of PRS and AGN in dwarf galaxies, it is likely that PRS have already been detected and perhaps misidentified as AGN \citep{2019MNRAS.488..685M}.

\begin{acknowledgements}
We acknowledge helpful discussions with Wenbin Lu, Kazumi Kashiyama, and Vikram Ravi.
CJL acknowledges support from the National Science Foundation under Grant No.\ 2022546. KA acknowledges support from NSF grants AAG-1714897 and \#2108673. 
\end{acknowledgements}

\bibliography{prsfrac}{}
\bibliographystyle{aasjournal}

\end{document}